\documentclass[man,floatsintext]{apa7}

\usepackage[american]{babel}
\usepackage{listings}
\usepackage{booktabs}
\usepackage{todonotes}
\definecolor{mygreen}{rgb}{0,0.6,0}

\usepackage[ruled]{algorithm2e} 
\usepackage{booktabs}

\graphicspath{{Fig/}} 
\usepackage{listings}
\definecolor{mygreen}{rgb}{0,0.6,0}

\usepackage{csquotes}
\usepackage[style=apa,sortcites=true,sorting=nyt,backend=biber]{biblatex}
\usepackage{mathtools}
\usepackage{amsthm,amssymb}
\usepackage{calrsfs}
\allowdisplaybreaks
\usepackage{hyperref}
\hypersetup{
  colorlinks   = true, 
  urlcolor     = red, 
  linkcolor    = blue, 
  citecolor   = blue 
}

\DeclareLanguageMapping{american}{american-apa}
\title{An information-theoretic approach to hypergraph psychometrics}
\shorttitle{Info-hypergraphs in psychometrics}
\authorsnames[1,1,{2,3,4,5},6,7,8,9,1]{Daniele Marinazzo,Jan Van Roozendaal,Fernando E. Rosas, Massimo Stella,Renzo Comolatti, Nigel Colenbier, Sebastiano Stramaglia, Yves Rosseel}
\authorsaffiliations{{Department of Data Analysis, Ghent University, Belgium},{Data Science Institute, Imperial College London, London, UK}, {Centre for Psychedelic Research, Imperial College London, London, UK}, {Centre for Complexity Science, Imperial College London, London, UK}, {Department of Informatics, University of Sussex, Brighton, UK}, {CogNosco Lab, Dept. of Computer Science, University of Exeter, UK}, {Department of Biomedical and Clinical Sciences “L. Sacco”, Università degli Studi di Milano, Italy}, {IRCCS San Camillo Hospital, Via Alberoni 70, Venice 30126, Italy}, {Physics Department, Universit\'{a} degli Studi di Bari Aldo Moro, and INFN Sezione di Bari, Italy}} 

\leftheader{Marinazzo}

\abstract{Psychological network approaches propose to see symptoms or questionnaire items as interconnected nodes, with links between them reflecting pairwise statistical dependencies evaluated on cross-sectional, time-series, or panel data. 
These networks constitute an established methodology to assess the interactions and relative importance of nodes/indicators, providing an important complement to other approaches such as factor analysis. However, focusing the modelling solely on pairwise relationships can neglect potentially critical information shared by groups of three or more variables in the form of higher-order interdependencies. To overcome this important limitation, here we propose an information-theoretic framework based on hypergraphs as psychometric models. As edges in hypergraphs are capable of encompassing several nodes together, this extension can thus provide a richer representation of the interactions that may exist among sets of psychological variables. 
Our results show how psychometric hypergraphs can highlight meaningful redundant and synergistic interactions on either simulated or state-of-art, re-analyzed psychometric datasets. 
Overall, our framework extends current network approaches while leading to new ways of assessing the data that differ at their core from other methods, extending the psychometric toolbox and opening promising avenues for future investigation.
}


\authornote{
\addORCIDlink{Daniele Marinazzo}{0000-0002-9803-0122}\\
\addORCIDlink{Fernando E. Rosas}{0000-0001-7790-6183}\\
\addORCIDlink{Massimo Stella}{0000-0003-1810-9699} \\
\addORCIDlink{Renzo Comolatti}{0000-0003-1097-3095} \\
\addORCIDlink{Nigel Colenbier}{0000-0003-0928-2668} \\
\addORCIDlink{Sebastiano Stramaglia}{0000-0002-5873-8564} \\
\addORCIDlink{Yves Rosseel}{0000-0002-4129-4477} \\

\vspace{2cm}
\noindent 
Correspondence concerning this article should be addressed to Daniele Marinazzo, Department of Data Analysis, Faculty of Psychological and Educational Sciences, Ghent University, 1 Henri Dunantlaan, B-9000 Ghent, Belgium.
\noindent
E-mail: \texttt{daniele.marinazzo@ugent.be}}

\begin{document}

\maketitle
\section{Introduction}
The last decade has seen the growth of various network approaches that allow researchers to probe and exploit potential interactions of observable variables associated with a given construct, or between constructs themselves~\parencite{schmittmann2013deconstructing, borsboom2013network,van2021bridges,marsman2018introduction,golino2017exploratory,borsboom2021network}. 
These approaches build networks where nodes represent measurable variables and links reflect relationships between them, 
with the purpose of focusing not on common underlying factors but on the structure of the inter-dependencies between the observables. Recently some approaches have been developed to include also factors in the network~\parencite{haslbeck2021moderated}, see below for further details. Such \textit{psychometric networks} are capable of reflecting dynamical processes~\parencite{schmittmann2013deconstructing}, and can  consider lagged relationships that can reveal dynamics leading to the emergence of symptoms~\parencite{borsboom2013network}.
Crucially, these networks often reveal non-trivial features of the groups of variables they represent, which can be identified by leveraging tools of network science such as measures of centrality ~\parencite{bringmann2019centrality,epskamp2018estimating} or community detection ~\parencite{van2021bridges}. Furthermore, it has been argued that important outgoing effects or events can be identified from an examination of either the overall network or part of it --- e.g. how interactions between certain symptoms may create a vicious circle triggering a series of negative events along the way~\parencite{schmittmann2013deconstructing}.

An illustrative case of the potential of network approaches can be found in the work on \textit{comorbidity}~\parencite{fried2017mental}, i.e. constructs that share multiple observables (e.g. depression and anxiety, which are sometimes considered as different facets of neuroticism
~\parencite{lovibond1995structure,van2021bridges}), and can be studied in terms of common edges or nodes, without limitations coming from modelling these sets via multiple latent variables
~\parencite{cramer2010comorbidity,jones2019bridge}. 

While network properties such as centrality and modularity provide valuable information about the overall topology of a network~\parencite{borsboom2021network}, traditional network approaches are limited in a fundamental yet often unacknowledged way: as their construction is based solely on pairwise links, they cannot give account of potential higher-order (i.e. beyond pairwise) interactions that involve three or more observables~\parencite{mcgill1954multivariate,rosas2016understanding,james2017multivariate}. In effect, the interactions between three or more variables are often nontrivial, poorly understood and, yet, can play fundamental roles in driving complex systems~\parencite{battiston2021physics}. 
From a statistical perspective, a pairwise approach assessing the relation between two variables (conditioned on other variables or not) 
cannot inform on the existence of synergistic or redundant interactions, which by definition involve three or more variables~\parencite{mcgill1954multivariate,williams2010nonnegative}.
Recent work is revealing how high-order interdependencies play key roles in diverse hallmark aspects of complex systems, including self-organisation~\parencite{rosas2018information}, synchronisation~\parencite{skardal2020higher}, 
percolation~\parencite{iacopini2019simplicial}, 
and emergent phenomena in general~\parencite{rosas2020reconciling}.

\begin{figure}[!ht]%
\includegraphics[width=.8\textwidth]{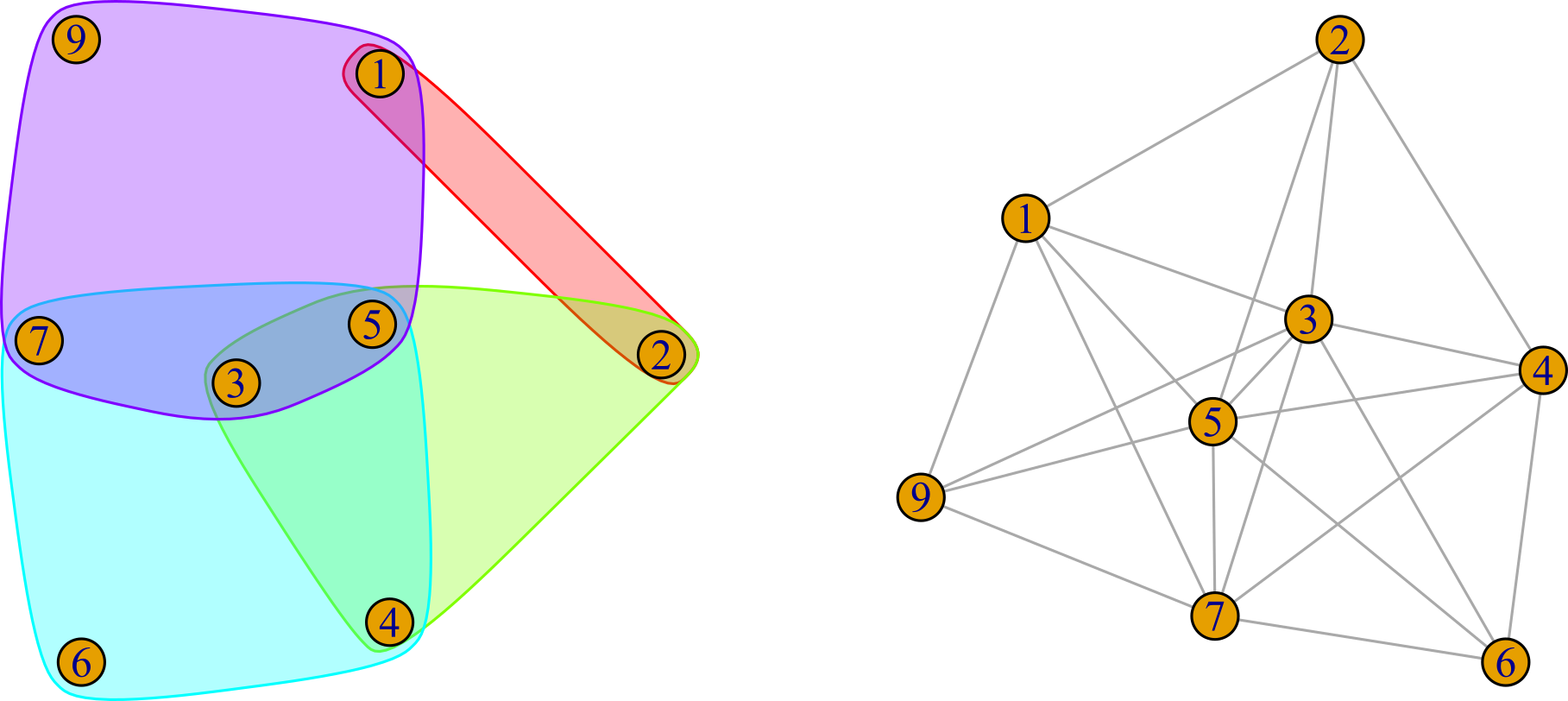}%
\centering
\caption{Illustrative example of an hypergraph with 8 nodes and 4 hyperedges (left) and its binarized version (right), generated with the R package \textit{HyperG}~\parencite{hyperg}.}%
\label{hypergraph_example}%
\end{figure}

Hypergraphs are the natural extension of networks that can properly give account of high-order interdependencies. 
In effect, analogously as graphs, hypergraphs are composed of nodes and \textit{hyperedges} (or hyperlinks) connecting them, with the key difference that hyperedges can connect more than two nodes nodes~\parencite{battiston2020networks}. Hyperedges can represent e.g. co-authors~\parencite{milojevic2014principles}, components of an ecosystem~\parencite{bairey2016high}, or individuals sharing a social or physical space in contagion models~\parencite{st2022influential}.
As an example, Figure~\ref{hypergraph_example} depicts an hypergraph and its binarized counterpart.

In the present work we put forward a psychometric framework that harnesses the power of hypergraphs to represent high-order interdependencies. 
At its core, our proposal is based on idea of 
moving from pairwise to a higher-order interdependencies as described in the multivariate joint distribution of the variables in question. 
As such, our proposed framework can be conceived as an extension of existent psychometric network approaches that, in turn, enables a systematic investigation of the role of redundant and synergistic interactions in psychometric scenarios 
to better characterise complex 
emergent behavior~\parencite{rosas2022disentangling}. In doing this, our approach is not necessarily overlapping, but rather complementary to the discovery of underlying structure such as latent factors and interactions, even in a \textit{higher-order} way as described in ~\parencite{haslbeck2021moderated}.
Our proposed framework is based on the \textit{O-information}~\parencite{Rosas2019}, an information-theoretic metric capable of individuating the interdependencies within multiplets of three and more variables 
(i.e. hyperlinks of statistical dependencies in the graph theory terminology), determining whether this common information is predominantly redundant or synergistic based on the sign of the informational content. 
The proposed framework is illustrated on synthetic datasets where the ground truth is known, and on standard datasets currently analyzed in network psychometrics studies.

\section{Methods}

\subsection{The state of the art: Exploratory Data Analysis and Networks}

Let us consider $n$ measurable variables $X_{1}, \ldots , X_{n}$, which could correspond to symptoms of a condition of interest, or items in a questionnaire. 
An attractive way to describe the interdependencies between these variables is to represent them as a  network~\parencite{vasiliauskaite2020understanding}. In general, networks are defined by two elements: (a) a set of nodes/vertices denoted by $V$, and (b) a collection of links/edges between the nodes denoted by $E$. Individual edges are usually denoted as $\textit{e}_{ij}$, where $i$ and $j$ represent the indices of two nodes. A network of observables typically represents each variable as a node, and uses links to depict relationships between the variables. 

Among the simplest and most used ways for establishing a relationship between two random variables ones finds (a) the Pearson correlation ~\parencite{mcgill1954multivariate}, and (b) the partial Pearson correlation, conditioned against all other considered variables~\parencite{van2021bridges}. However, both these approaches are affected by important limitations. In effect, it has been convincingly shown that using plain correlations to build network links has serious drawbacks in terms of potentially highlighting spurious correlations~\parencite{epskamp2018tutorial,bhushan2019using}. While this issue can be partially avoided by adopting conditional correlations, these in turn can also be problematic in the presence of colliders~\parencite{rohrer2018thinking} and in general of shared informational content~\parencite{stramaglia2014synergy}. Specifically, network analyses typically consider networks whose links are given by the \textit{precision matrix} $C = \textstyle\sum^{-1}$ with $\textstyle\sum$ being the correlation matrix, which allows to calculate partial correlations via $\rho_{ij}=\frac{C_{ij}}{\sqrt(C_{ii} \dot C_{jj})}$. It is worth mentioning that accurate estimations of the precision matrix are non-trivial and prone to biases, but adequate techniques have been developed in recent years~\parencite{epskamp2018estimating}. In particular, it is common to use regularization techniques such as LASSO ~\parencite{golino2017exploratory,golino2021modeling} to restrict the number of non-zero entries in the precision matrix.

\subsection{Measuring higher-order effects via the $O$-information}

While Shannon's mutual information~\parencite{shannon1948mathematical} may provide an encompassing account on the interdependencies between two (sets of) variables, it is unable to fully explore the rich interplay that can take place within triple or higher-order interactions~\parencite{mcgill1954multivariate,rosas2016understanding,james2017multivariate}. Two popular multivariate extensions of the Shannon mutual information are the \textit{Total Correlation} (TC)~\parencite{watanabe_information_1960} and the \textit{Dual Total Correlation} (DTC)~\parencite{te_sun_nonnegative_1978}, which can be calculated as
\begin{align}
\text{TC}(X_1,\dots,X_n) &:= I(X_1;X_2) + I(\bm X_1^2; X_3) + \dots + I(\bm X_1^{n-1};X_n) , \label{eq:TC}\\
\text{DTC}(X_1,\dots,X_n) &:= I(X_1;X_2|\bm X_3^n) + I(\bm X_1^2;X_3|\bm X_4^n) + \dots + I(\bm X_1^{n-1};X_n) \label{eq:DTC}
\end{align}
Above, $I(X;Y)$ corresponds to the Shannon mutual information between $X$ and $Y$, $I(X;Y|Z)$ to the conditional mutual information given $Z$, and $\bm X_i^j=(X_i,\dots,X_j)$ is a shorthand notation. 
Both the TC and DTC are non-negative quantities that are zero if and only if all variables $X_1, \dots, X_N$ are jointly statistically independent --- i.e. if $p(X_1,\dots,X_n) = \prod_{i=1}^n p(X_i)$. 
Interestingly, both are equal to the mutual information for $n=2$ but differ when $n\geq 3$. 
In fact, both metrics provide distinct but complementary views on multivariate interdependencies: it can be shown that $\text{TC}$ accounts for the effect of \emph{collective constraints} (i.e. regions of the phase space that the system explores less in its dynamics), while the $\text{DTC}$ measures the amount of \emph{shared randomness} between the variables (i.e. the information that can be collected in one variable that also refers to the activity of another)~\parencite{Rosas2019}.

An attractive way to exploit these complementary views is by considering their difference,
\begin{align}
  \Omega_n(X_1,\dots,X_n) &= \text{TC}(X_1,\dots,X_n) - \text{DTC}(X_1,\dots,X_n)~. 
  \label{eq:Oinfo}
\end{align}
which is known as the \emph{O-information}~\parencite{Rosas2019}. 
The O-information is a signed metric that captures the balance between high- and low-order statistical constraints: Low-order constraints impose strong restrictions on the system and allow little amount of shared information between variables, while high-order constraints impose collective restrictions that enable large amounts of shared randomness. By construction, $\Omega (X_1,\dots,X_n) < 0$ implies a predominance of high-order constraints within the system of interest, a condition that is usually referred to as \textit{statistical synergy}. Conversely, $\Omega (X_1,\dots,X_n) > 0$ implies that the system is dominated by low-order constraints, which imply \textit{redundancy} of information. 
For more details related to the O-information and its interpretation, we refer the reader to~\parencite{Rosas2019}.

\subsection{Measuring higher-order effects on data}

Let us consider a scenario where we have data from $n$ variables of interest, denoted by $X_1,\dots,X_n$. It is important to note that the O-information of the whole, $\Omega(X_1,\dots,X_n)$, only provides information about the \textit{dominant trend}, and hence it is also interesting to calculate the O-information of \textit{multiplets}, i.e. subsets of variables, as they can show a different dominant trend that may be buried among other patterns. 

Following these considerations, we employ the following analytical pipeline:

\begin{enumerate}
    \item Calculate the $O$-information for multiplets of sizes from $3$ to $m$ (with $m\leq n$ chosen considering computational convenience in case $n$ is too large, or specific questions; the multiplet with n=2 represents pairwise mutual information).
    \item Partition the multiplets depending on the sign of the O-information into positive (redundant) and negative (synergistic).
    \item Evaluate the significance of the O-information values via bootstrap  ~\parencite{efron1994introduction}.
    \item Compare the bootstrap confidence interval~\parencite{diciccio1996bootstrap} of each multiplet with subsets of its components one order below. Multiplets are discarded if the confidence intervals overlap, as this indicates that they do not display additional synergy or redundancy.
    \item Build two hypergraphs with the significant multiplets: one with hyperedges of positive O-information related to dominantly redundant interdependencies, and another with hyperedges of negative O-inforamtion associated to synergy.
\end{enumerate}
The output of this pipeline are \textit{redundancy} and a \textit{synergy hypergraphs}, which give an overview of the structure of interdependecies in the system by encoding the multiplets whose interaction are dominantly redundant or synergistic, respectively. These two hypergraphs can then be used to perform further analysis related to their topology, including centrality or clustering analyses.

In this paper, step 1 is carried out via a Gaussian copula estimator, as described in~\parencite{ince2017}, which allows a parsimonious yet robust estimation of the O-information. 
For step 3, the significance is evaluated by means of a bootstrap procedure. The significant multiplets are those with a confidence interval not containing zero and have an associated $p$-value smaller than a fixed significance level, e.g. $0.01$), corrected for multiple comparisons using the Bonferroni-Holm false discovery rate (FDR). 
After step 4 is completed, the output will be a collection $C$ of multiplets of different lengths, ${X_1,\dots,X_m}$ with $m < n$, featuring any of the $n$ variables under analysis. This collection $C$ can then be interpreted as a list of hyperlinks indicating non-overlapping higher order interactions between the considered variables. For illustration, the analyses presented in this paper focus on multiplets of order 3, 4, and 5.

\subsection{Generating higher-order effects on synthetic data}

In order to test hypotheses pertaining to higher order properties, it is useful to have the capability to generate synthetic data with specific levels of higher order information. For this purpose, we employ a general approach based on three steps, which are 
described as follows. First, we construct triplets of variables generically named $\textit{x,y,z}$ with a pre-specified level of higher order information --- 
specifically, we focus on multivariate Gaussian systems and  
build $3 \times 3$ covariance matrices that exhibits a desired level of higher
order information. These covariance matrices are then stacked on the diagonal 
of a $P \times P$ covariance matrix, where $P$ is the
total number of nodes in the system. As second step, we build connections between some of these triplets (if desired) by setting the corresponding elements of the off-diagonal covariance matrix to non-zero values. 
As final step, we use the resulting covariance matrix to generate multivariate Gaussian data. To reduce the corresponding degrees of freedom,
we set the means to zero; moreover as the scale is not important we 
standardize each of the variables --- so that the covariance matrix
of the resulting data is in fact a correlation matrix. The most challenging step of this process is the construction of $3 \times 3$ correlation matrices with a pre-specified level of higher order information --- for which we developed a numerical pipeline using the R package \textit{lavaan} ~\parencite{rosseel2012lavaan}, which is described in Appendix~\ref{app:procedure}.

\begin{figure}[!ht]%
\includegraphics[width=.8\textwidth]{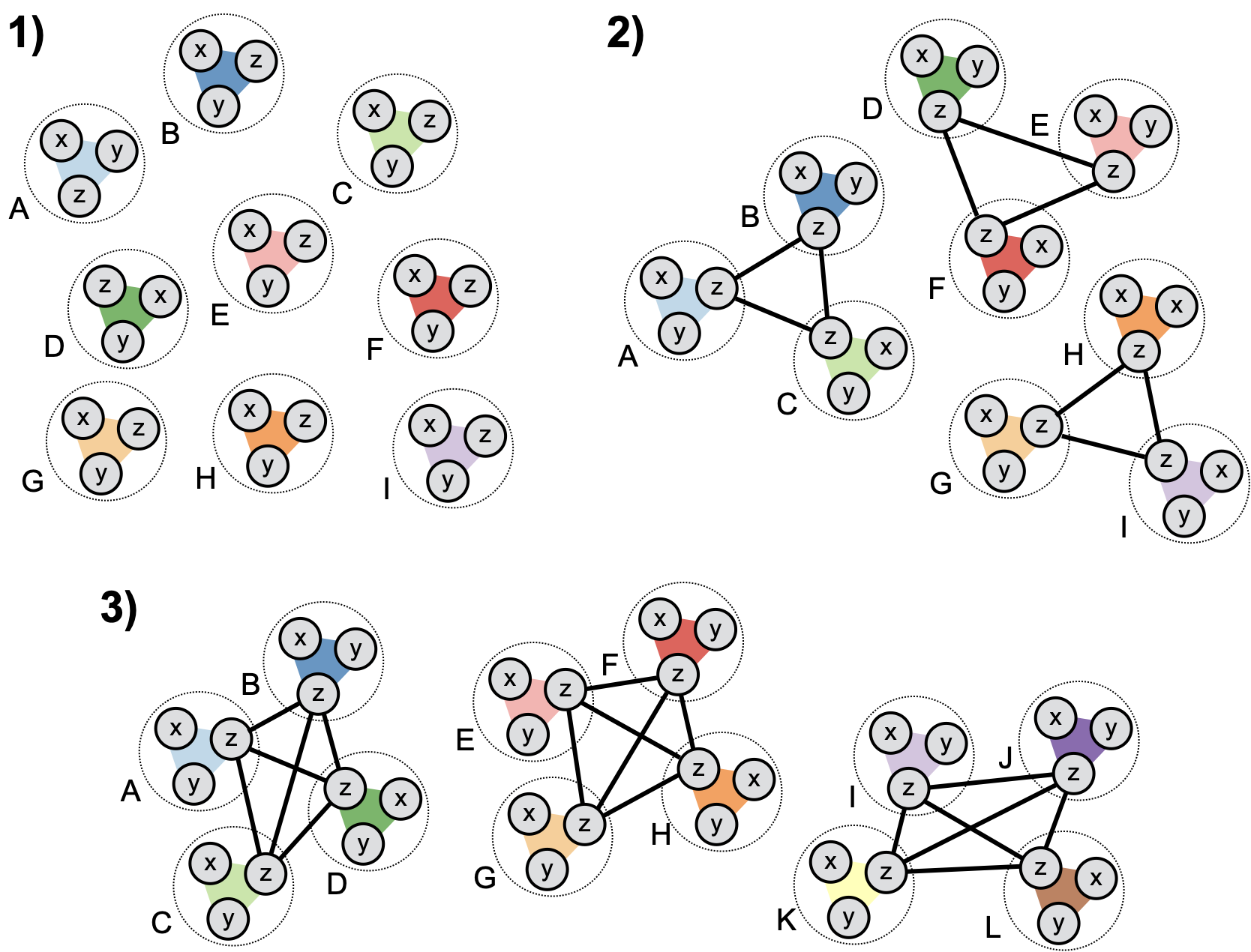}%
\centering
\caption{Visual overview of the presented model layouts. 
\textbf{1)} Nine independent triplets. \textbf{2)} Three clusters of three triplets. All triplets within a cluster are connected to two others via their \textit{z}-variables. 
\textbf{3)} Expansion of Model Layout 2. Three clusters of four triplets. Each triplet within a cluster is connected to the three other related triplets via the \textit{z}-variables.}
\label{model_collection}%
\end{figure}

Having discussed the basic elements of our procedure, we now present several potential model variations, changing the level of interaction information for one or more triplets. For this purpose, each of the suggested models has been scaled up to include between nine and twelve triplets. All of the following layouts will use one particular set of residual correlations, meaning that only residual correlations are used between \textit{z}-variables, or those only between \textit{y}- and \textit{z}-variables. The layout designs in figure \ref{model_collection} show how each node from a triplet can be connected to two, three or four nodes originating from other triplets.

\newpage

The levels of the residual correlations per model correspond to the maximum allowed in the case of forcing each triplet to include the maximum possible amount of synergy within the model.
\newpage

\section{Results}

\subsection{Synthetic datasets}

\begin{figure}[!htb]%
\includegraphics[width=\textwidth]{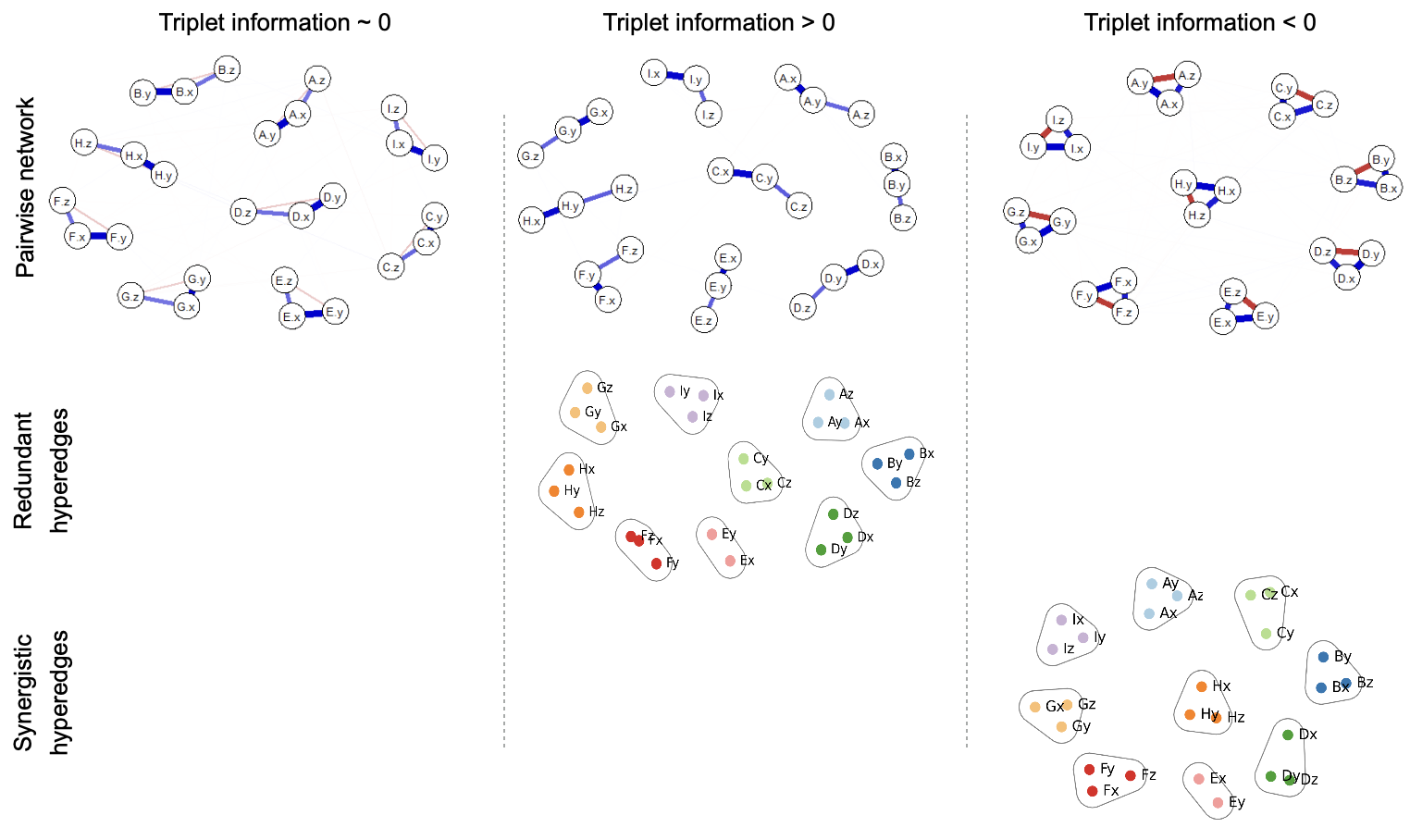}%
\centering
\caption{Results obtained from Model 1 in Fig.~\ref{model_collection} when implemented with zero (i.e. independent), positive (i.e. redundant), and negative (i.e. synergistic) interaction information in each triplet. \textbf{Top row:} network obtained with $bootnet$~\parencite{epskamp2018estimating} and represented with $qgraph$~\parencite{epskamp2012qgraph}, where which both color intensity and edge thickness represent edge weight. Red indicates negative edge weight,
green indicates positive edge weight. Left: synergy; Middle: Zero interaction
information; Right: redundancy. \textbf{Middle row:} significant multiplets of redundant $O$-information. \textbf{Bottom row:} significant multiplets of synergistic $O$-information.}%
\label{fig_model1}%
\end{figure}

We first analysed a system composed of 27 variables arranged into nine triplets with various covariance structures, which were --- as triplets --- independent of each other. We considered three covariance structures, which give zero, positive, and negative O-information, respectively (we informally refer to the O-information of these triplets as ``triplet information''). As expected, our analysis pipeline correctly identifies redundant and synergistic interdependencies in multiplets of order 3, but nothing in higher orders \ref{fig_model1}. This provided a confirmation that our analysis pipeline is accurate in this simple scenario.

\begin{figure}[!htb]%
\includegraphics[width=\textwidth]{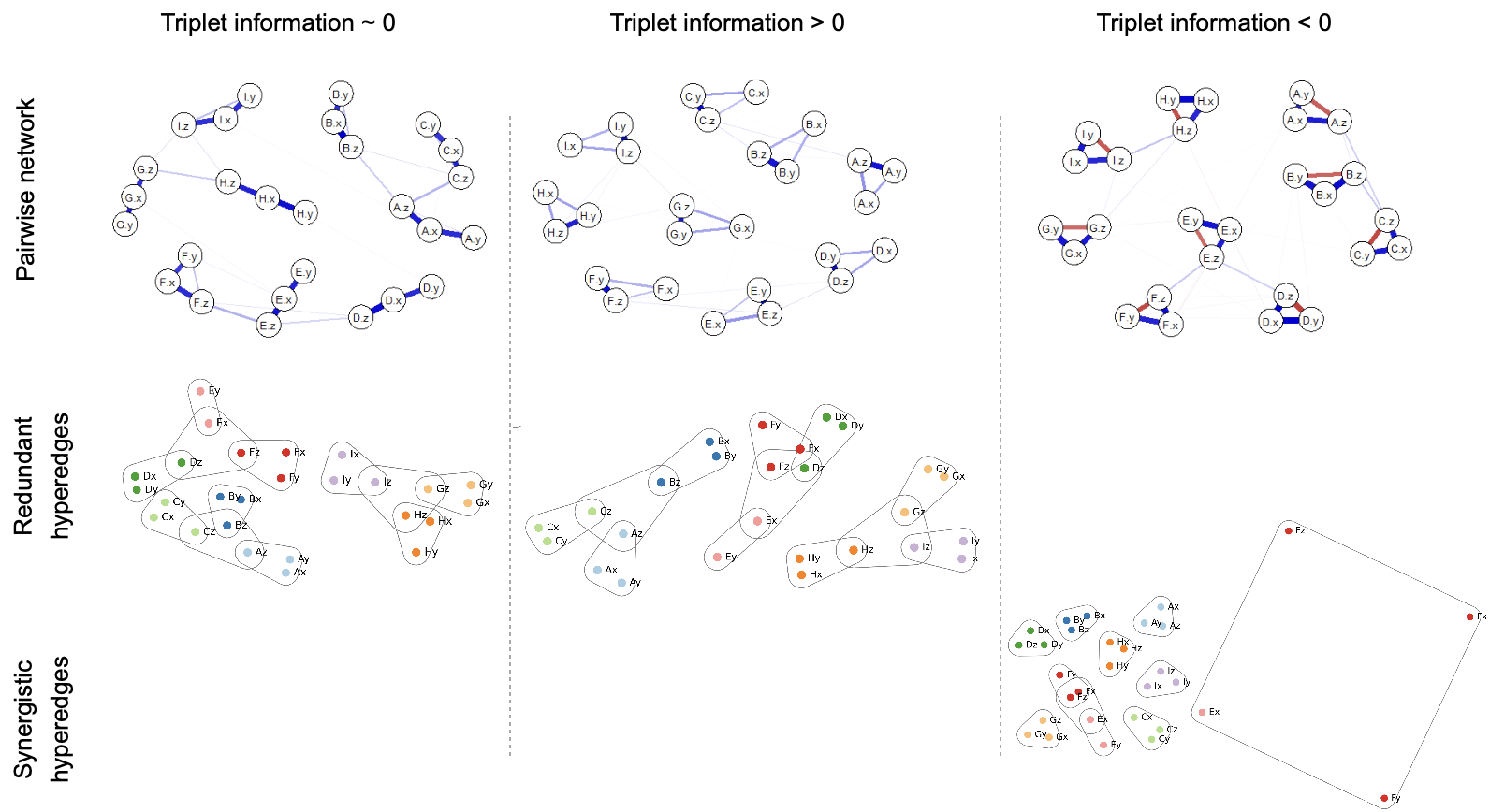}%
\centering
\caption{Analogous results as shown in Figure~\ref{fig_model1} but for Model 2 in Figure \ref{model_collection}.}
\label{fig_model2}%
\end{figure}

As a second step, we analysed scenarios where three groups of three triplets each were linked through their \textit{z} variables (see Figure~\ref{model_collection}). 
These links through the $z$ variables in each triplet --- weaker than the inter-triplet links --- resulted in the appearance of redundant multiplets in the independent case \ref{fig_model2}. For the case of positive triplet information, these links result on new redundant triplets connecting the $z$ variables of different triplets, generating a more connected hypernetwork. The synergistic triplets are not much disturbed by these links, while one multiplet of order 4 appears. Interestingly, while the resulting hypergraph of redundant hyperedges follows the pairwise structure, it is much less straightforward to derive the resulting synergistic hyperedges.

%

Finally, we analysed scenarios with three groups of four triplets linked through the \textit{z} variables. 
The between-triplets links result in synergistic multiplets of order 3 also in the independent and redundant cases. In the synergistic case synergy extends to order 4 maintaining the same integration of linked multiplets. Overall, the inclusion of larger cliques result in a richer array of higher order interdependencies, resulting in redundant and synergistic hypergraphs that depart more dramatically from the corresponding pairwise network.
\begin{figure}[!htb]%
\includegraphics[width=\textwidth]{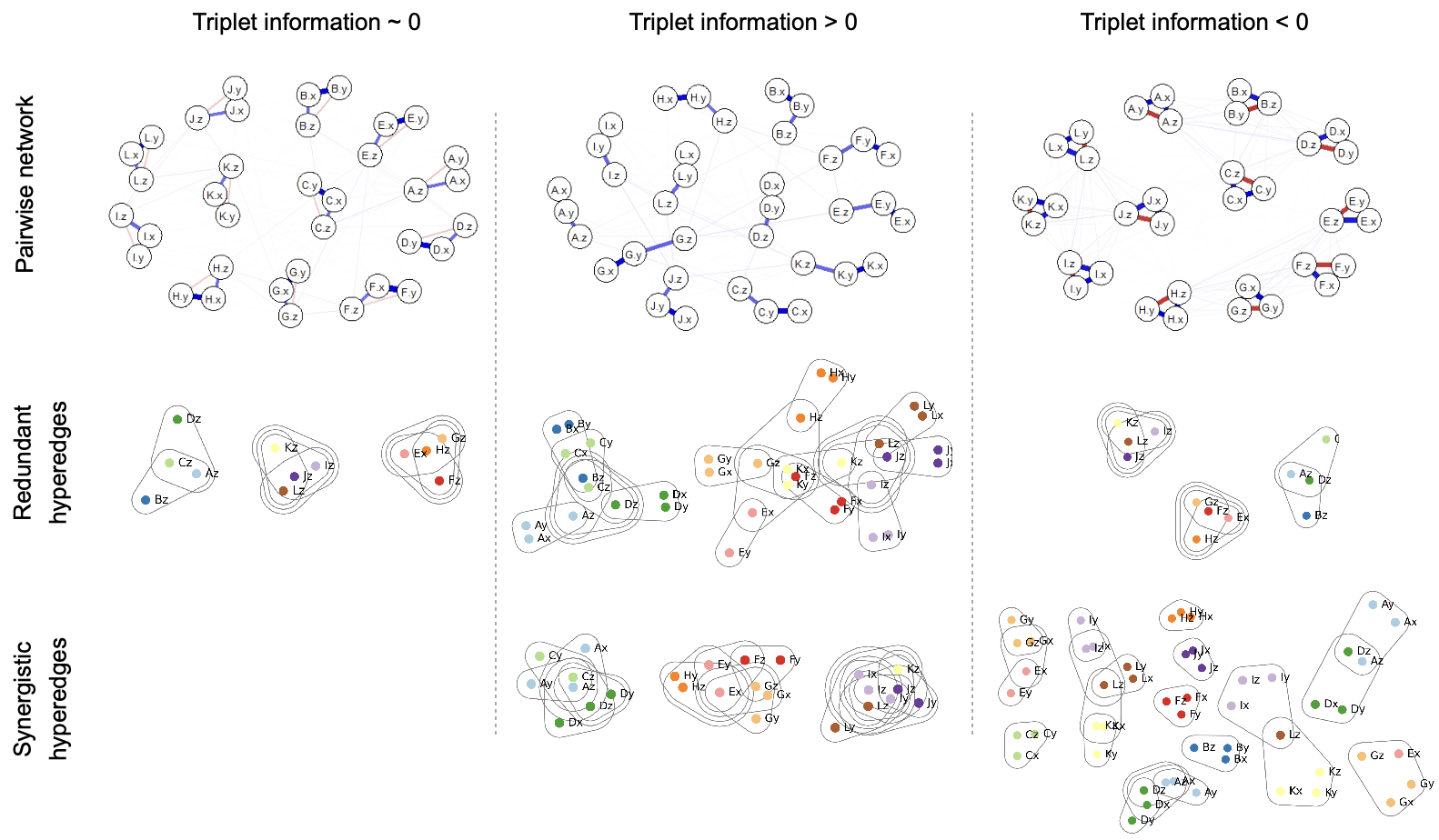}%
\centering
\caption{Analogous results as shown in Figure~\ref{fig_model1} but for Model 3 in Figure \ref{model_collection}.}
\label{fig_model4}%
\end{figure}

\subsection{Reanalysis of the Empathy dataset ---~\parencite{briganti2018network}}

\begin{figure}[!ht]%
\includegraphics[width=\textwidth]{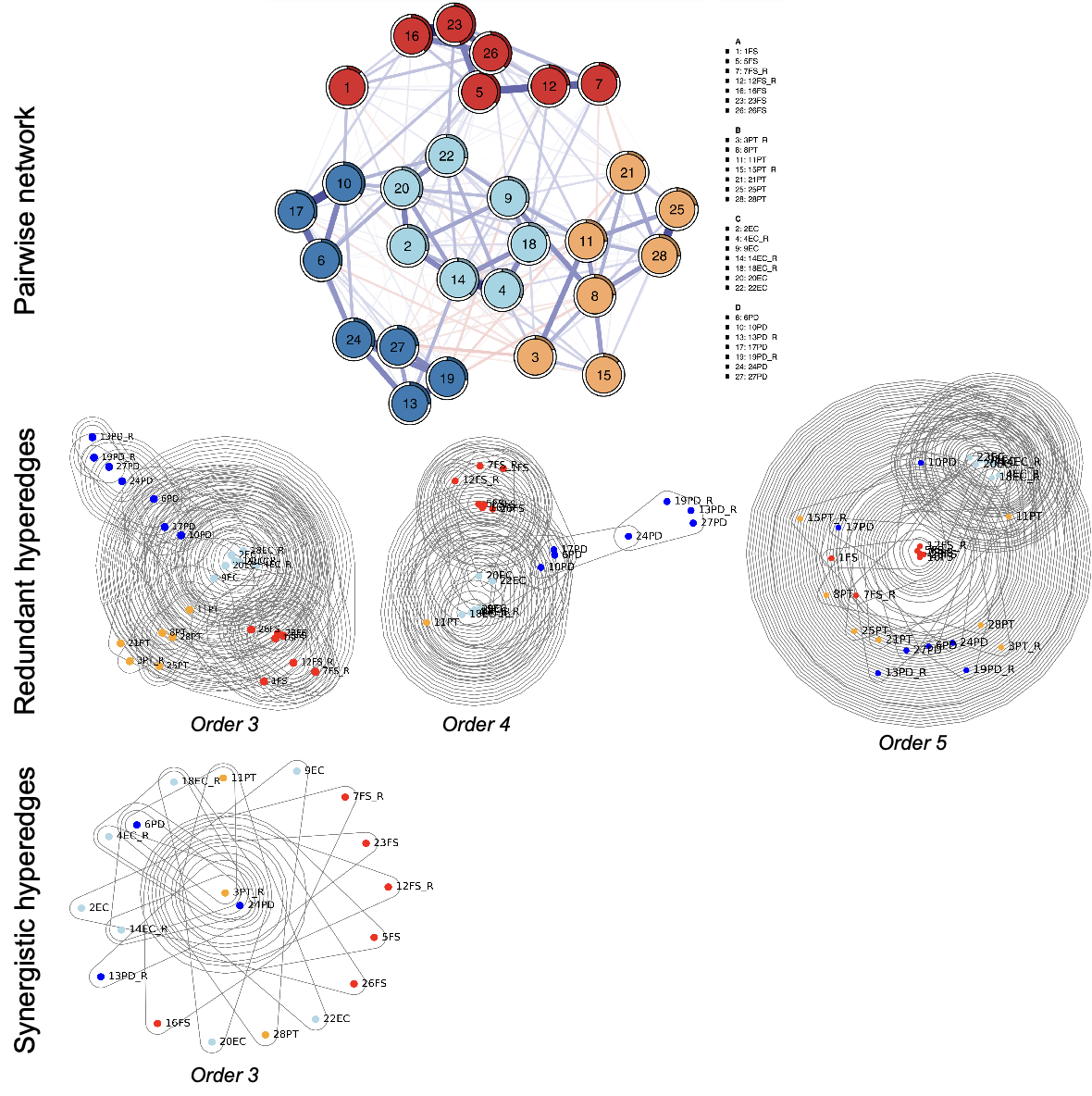}%
\centering
\caption{Empathy dataset from \protect\parencite{briganti2018network}. Colors indicate membership of a component of items, as in the reference paper. Top row: network obtained with $bootnet$. Middle row: significant multiplets of redundant $O$-information. Bottom row: significant multiplets of synergistic $O$-information.}%
\label{fig_empathy}%
\end{figure}

We re-analysed the psychometric data presented in \cite{briganti2018network} around the construct of empathy ($N=1973$). We find (\ref{fig_empathy}) that the O-information can identify significant synergistic and redundant interactions co-existing within the hypergraph representation of the construct. Interestingly, multiplets of orders 3 and 4 identify the same major factors obtained with the pairwise psychometric network construction --- namely fantasy, perspective taking, personal distress and empathic concern. Since our network embeddings tend to place at the center items engaging in more multiplets at once, our hypergraph visualisations confirm Briganti and colleagues' findings of empathic concern being a pivotal dimension of empathy. However, this happens only when redundant interactions up to order 3 and 4: at higher multiplet orders fantasy becomes a more relevant major factor. This highlights higher order interactions between cognitive-related projection abilities into fictional characters, measuring fantasy, and more emotionally-focused items relative to distress, empathy, and anticipation into the future. Please note that these distinct trends could not have been retrieved by pairwise psychometric networks. 

These results agree with recent independent studies indicating a stronger-than-expected interplay between fantasy and emotions in regulating empathy~\parencite{nomura2012empathy}. Interestingly, synergistic interactions highlight items 3 and 24 as pivotal. Whereas item 3 relates to the ability to see things from the perspective of other individuals, and is thus evidently crucial for synergistic interactions relative to empathy, item 24 (``I tend to lose control during emergencies'') is relative to emotional regulation and distress. Our results indicate that emotional regulation --- the ability to exert control over emotional responses --- provides information over the whole construct when combined with items of all other four factors. 

The retrieved informational hyperedges provide relevant quantitative support to other recent studies~\parencite{thompson2019empathy} showing how emotional regulation and empathy greatly overlap, mainly in terms of understanding emotions, although at different levels. Although the causal directionality between empathy and emotion regulation has not been clearly assessed, exploratory studies have shown how a lack of empathy can trigger emotional dysregulation in young adults~\parencite{schipper2013relating}, further confirming the synergistic centrality of emotion regulation found in our hypergraph representation.

\subsection{Reanalysis of the PTSD dataset ---~\parencite{ARMOUR201749}}

\begin{figure}[!ht]%
\includegraphics[width=12cm]{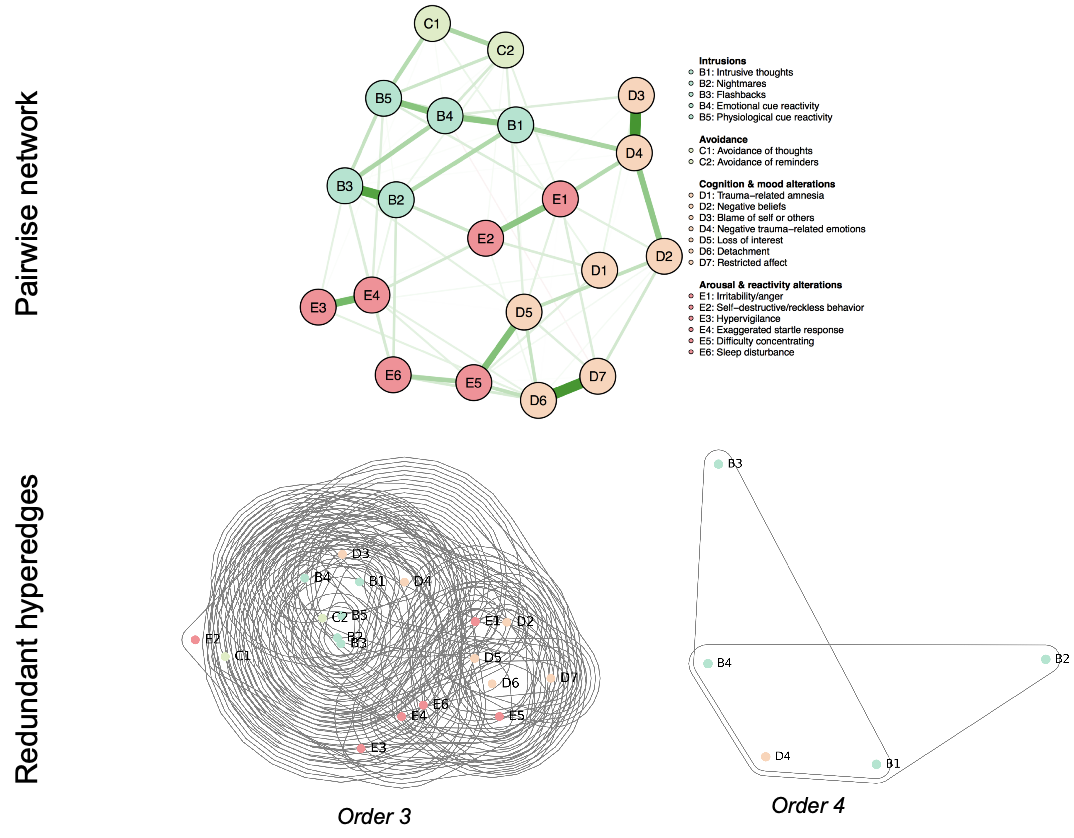}
\centering
\caption{PTSD dataset from \protect\parencite{ARMOUR201749}. Colors indicate membership of a class of items. Top row: network obtained with $bootnet$. Middle row: significant multiplets of redundant $O$-information. Bottom row: significant multiplets of synergistic $O$-information.}%
\label{fig_PTSD}%
\end{figure}

As a second re-analysis, we studied the PTSD dataset~\parencite{ARMOUR201749} which includes information about $N=221$, U.S. veterans. 
Our analyses did not identify significant synergistic interactions between variables (\ref{fig_PTSD}), which reflects a lack of high-order interdependencies across items in the four major factors --- which include elements as distinct as intrusions, avoidance, cognitive and mood alterations, and arousal and reactivity alterations. In contrast, several redundant interactions where found among items in multiplets of orders 3 and 4, which suggest that these factors may be driven by similar underlying factors. These redundant O-information hyperedges identifies items about nightmares, flashbacks, physiological cue reactivity and avoidance of reminders as being highly central. This combination of behavioural, emotional and memory-related patterns underlines the complexity of PTSD as a condition affecting not only the emotional and cognitive spheres of individuals, but also their behaviour and physiology~\parencite{armour2020posttraumatic}. 

\begin{figure}[!ht]%
\includegraphics[width=\textwidth]{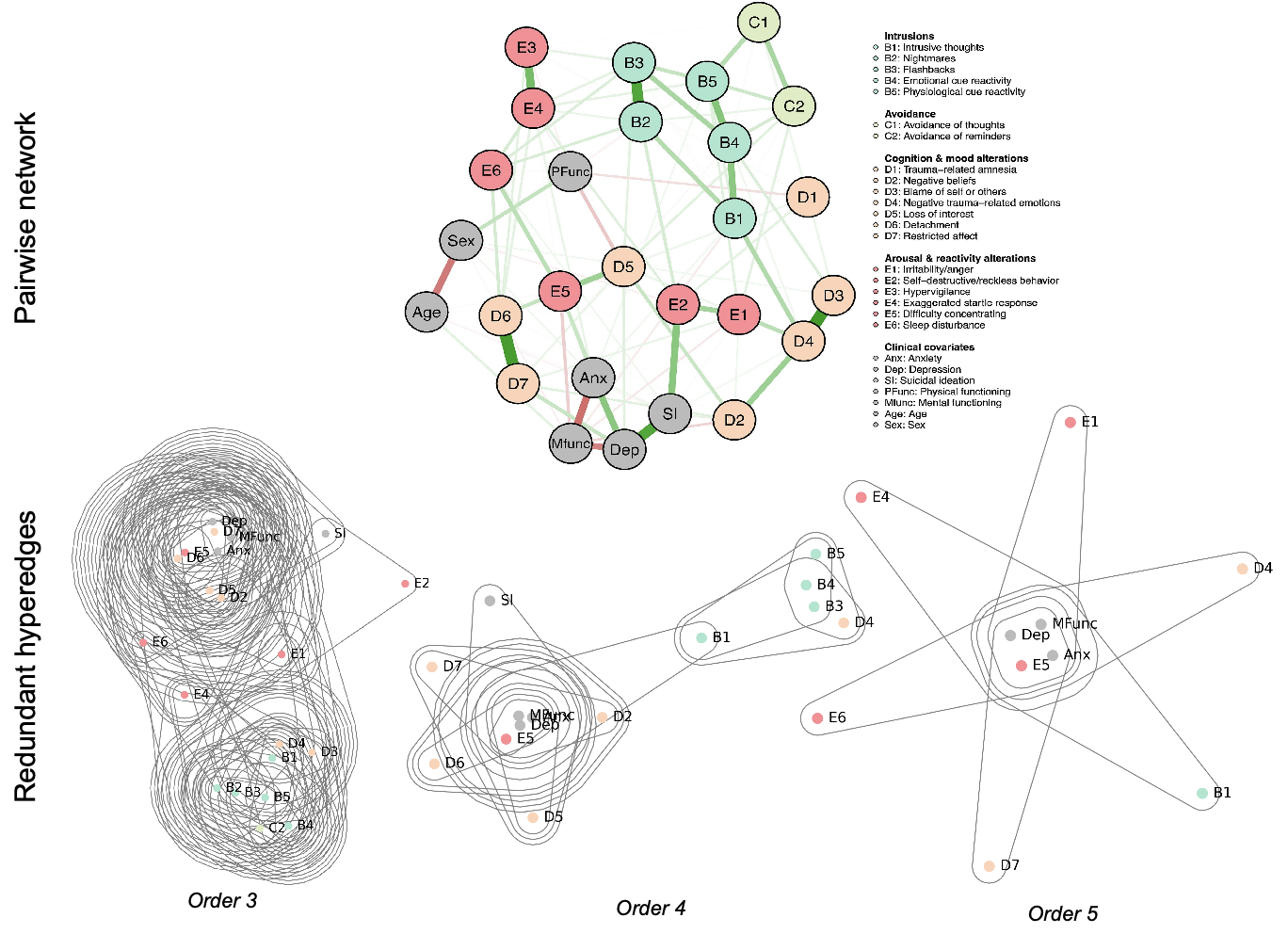}%
\centering
\caption{PTSD dataset from \protect\parencite{ARMOUR201749}. Colors indicate membership of a class of items. Grey items are the covariates. Top row: network obtained with $bootnet$. Middle row: significant multiplets of redundant $O$-information. Bottom row: significant multiplets of synergistic $O$-information.}%
\label{fig_PTSDfull}%
\end{figure}

The lack of a clearer internal structure and of synergistic interactions might be due not only to an intrinsic complexity of the construct, but also to sample size issues, missing item in scales and other methodological issues. 
To better understand this, we re-analysed the same dataset while using a larger number of variables~\parencite{ARMOUR201749} to include covariates (\ref{fig_PTSDfull}). In this second hypergraph, clinical covariates like anxiety and depression levels were identified as being pivotal for redundant interactions across multiplet orders 3, 4 and 5. This indicates that anxiety and depression levels overlap significantly with behavioural, emotional and physiological symptoms of PSTD, in agreement with independent studies~\parencite{armour2020posttraumatic}. The addition of clinical covariates polarised the hypergraph structure and led to a more pronounced separation between intrusion variables and others. Furthermore, intrusive thoughts were found to be an item bridging multiplets at order 4 in terms of redundant information. Hence, our analysis indicates a key role mediated by intrusive roles as possessing richly redundant information and thus being relevant for understanding the construct. 

Our findings agree with other studies highlighting intrusive thoughts as key signs of PTSD~\parencite{brooks2019trauma}, potentially rising from a maladaptive thought suppression, i.e. a conscious, imperfect, cognitive-driven suppression of a negative thought can enhance the prominence and recurrence of the thought itself, causing worry and rumination. The persistence of intrusive thoughts has serious documented consequences over behaviour and emotional wellbeing \cite{brooks2019trauma,armour2020posttraumatic}, which support the redundancy higher order patterns found in our work. Interestingly, synergistic interactions miss even in this expanded set of items. Future research should investigate whether this pattern is due to sample size issues or other methodological aspects of PTSD measurements.

\section{Discussion}

We have put forward a novel approach to capture higher order interactions between observables in psychometrics and described them in the form of hypergraphs. In particular, we introduced an analysis pipeline that builds two hypegraphs reflecting redundant and  synergistic interdependencies. This approach capitalizes on recent advances in multivariate information theory, which have provided data-efficient metrics to capture these phenomena in multiplets of different sizes, and has the scope to build up and extend existent network approaches to psychometric. 


Current network approaches to psychometric are based on a key assumption: that relationships between variables can most of the time be represented as pairwise networks --- either in cases of having a common cause or effect, or being linked by a directed flow of influences (i.e. as a directed acyclic graph) or a Ising-like structure~\parencite{borsboom2021network}. This assumption, in turn, leads to the adoption of pairwise random Markov fields as a natural modelling choice for identifying the interactions between variables. 
However, recent work~\parencite{williams2011information,mediano2021towards,grasso2021causal} has convincingly shown the relevance of higher order interdependencies, in which relationships often take place between groups of variables in a way that cannot be reduced to a set of pairwise links. 
Building on these insights, our proposed framework aims to relax these assumptions along two axes.

On the first axis, we propose to disentangle the analysis of high order mechanisms (and structures) from the high order behaviors that can emerge in complex systems. As described in~\parencite{rosas2022disentangling} in the context of physical systems, there is a fundamental distinction between mechanisms that address how the system is structured, and behaviors that correspond to emergent properties related to what the system ``does.'' Furthermore, the existence of higher order mechanisms do not imply high order behaviors or \textit{vice-versa}. Consequently, our proposed approach focuses on the analysis of high order patterns in the data, while explicitly acknowledging that this does not bring straightforward implications on how such data was generated in the first place. This angle also underlines the main difference with another way of seeing higher-order networks in psychometrics, where the links involve latent factors considered as mediators in three-way relationships evidenced by mixed graphical models~\parencite{haslbeck2021moderated} \footnote{To stress where this approach is different, consider a simple linear regression $Y \sim X1 + X2$. It could be that the triplet $(Y, X1, X2)$ exhibits redundancy, synergy or independence. According to~\parencite{haslbeck2021moderated}, a \textit{second-order} model would be $Y \sim X1 + X2 + X1*X2$, involving the addition of more information/variables (product terms). This \textit{second-order} model would need a 4x4 covariance matrix, while our approach would stick to a 3x3 covariance matrix. On the other hand one could, of course, create a quadruplet (Y, X1, X2, X1*X2) that exhibits redundancy or synergy. In other words, the two definitions are complementary, and they can be combined, but they refer to different phenomena.}.

The second axis relates to possible confounding effects, and their nature. Variables sharing information, yet considered as individual or pairwise interacting units, can pose problems for proper network reconstruction and in assigning a given centrality score to an individual variable. 
The presence of a variable could either enhance or deteriorate statistical dependencies between other variables, possibly influencing results for network estimation when using established centrality indices of strength, betweenness, and closeness
~\parencite{hallquist2021problems}. Analogous issues of deteriorated model validity would hold also for other approaches like factor analysis~\parencite{tuccitto2010internal}. Another possibility could be the existence of a statistical dependency between two variables which is considered insignificant in an exploratory analysis~\parencite{golino2017exploratory,epskamp2018estimating}. This issue increases the concern of detecting false positives regarding statistical dependencies during estimation. While some efforts have been done in this direction (e.g. 
~\parencite{haslbeck2021moderated}), 
correlations can persist even after conditioning on a latent variable, a phenomenon known as local dependency~\parencite{mcdonald2013test}. The persistence of correlated residuals can originate from multiple causes~\parencite{golino2021modeling,mcdonald2013test,rust2014modern,wang2005exploring}, e.g. by construction (scale items are selected as already being strongly correlated with each other), because of variables sharing semantic framing (in their item descriptions), or because of a lack of randomisation in the order of items and subsequent cross-item contextual assimilation effects. Local dependency violates the assumptions of commonly used psychometric models of latent variables, like Item Response Theory or factor analysis~\parencite{mcdonald2013test}. If significant correlations persist between items even after the contribution of the latent variable is removed, then items are locally dependent and this creates unaccounted model dimensions, with consequences like factor loading inflation or factor splitting in noisy/theoretically unexpected sets of variables. Advantageous solutions to overcome local independence are either to bundle sets of locally dependent items into polytomous super-items~\parencite{mcdonald2013test,rust2014modern} in the setting where the items are binary, or consider 3-way interactions~\parencite{haslbeck2021moderated}. Both these advancements pose novel challenges about preserving statistical, diagnostic and interpretable measures. Our work contributes quantitatively to this direction while exploiting the powerful statistical framework of higher order interactions in complex networks~\parencite{rosas2022disentangling}. Considering higher order interactions involving the observables only, and thus transferring these directly to hyperedges may let network approaches even more complementary to classical factor analysis.

Our approach may also be helpful for 
detecting latent factors and correctly attribute observables to them, which is also the goal of recent work aimed at modelling local dependency as a trace for \textit{redundancy}~\parencite{christensen2020unique,golino2021entropy,golino2021modeling,vijayakumar_unified_2022}. Redundancy comes from the fact that within a latent variable model, locally dependent variables contain less information than the one predicted by the (inflated) model parameters. Christensen and colleagues~\parencite{golino2021modeling} measured redundancy and local dependency by considering psychometric networks of pairwise relationships enriched with weighted topological overlap, i.e. a metric estimating how similarly connected are any two nodes in a network. Their innovative approach showed how considering ``higher order'' interactions through network overlap provided superior factor estimation compared to IRT modelling. Our approach adopts a different perspective, substituting the pairwise psychometric network with a hypergraph structure in which  higher order interactions are naturally captured by information-theoretic hyperlinks. Thus, it is worth noting that the terms ``higher order,'' ``redundancy'' or ``3-way interactions'' can have a different flavor or even some different definitions than the ones used so far. An important difference is that the structures simulated in the papers cited above do indeed generate redundancy (in the sense of high order informational content), but not synergy (see \ref{app:Golino_red}). Whereas redundancy can be captured by local dependency between variables, synergy must be interpreted in different terms~\parencite{stramaglia2016synergetic}, i.e. as the amount of information available only when variables are jointly considered. Our approach enables the quantification of synergistic interactions also within psychometric networks, extending to psychological representations a measure originally devised for brain data and networks.


The presented framework uses the O-information in its basic form, but there is a number of variations of it which can be used for future work. In effect, recent work have introduced multiple extensions of the O-information, including dynamical versions~\parencite{stramaglia2021quantifying} that provide an expansion of Transfer Entropy (or Granger Causality in the Gaussian case~\parencite{barnett2009granger}), point-wise extensions which provide O-information values for each individual pattern~\parencite{scagliarini2022quantifying}, and a spectral decomposition that resolve the O-information into different frequency bands~\parencite{faes2022framework}. 
These tools could be used to build hypergraphs extending temporal psychological networks, as those presented in~\parencite{fried2022mental}. 
Please note that the metrics introduced in~\parencite{stramaglia2021quantifying} can be used to calculate conditional O-information metrics, which provide a close simile to the conditional correlation that is often used in the network psychometric literature. 
Other interesting avenues of interesting future work include leveraging recent algorithms to identify hypergraph centrality~\parencite{tudisco2021node} and modularity~\parencite{kaminski2020community,chodrow2021generative}, and also harmonic analysis and dimensionality reduction~\parencite{medina2021hyperharmonic}.

One limitation of the O-information (and any other coarse-grained metric over an information decomposition~\parencite{williams2010nonnegative})  
is that a multiplet is labelled as synergistic or redundant depending on its sign, and in this sense it reflects the net balance between the two quantities and only acknowledges the dominant one. Note that redundancy might originate from underlying psychological correlations, 
for instance two or more symptoms might overlap rather than causing one another, thus bringing more redundancy within the analysis~\parencite{epskamp2018estimating}. For these reasons redundancy is prevalent in real datasets --- where it is most likely to be found by experimental design or construction, and hence this could cover synergistic interactions making them harder to find. 
One solution could be to first reduce the redundancy in the sense of common factors~\parencite{golino2021modeling}, or consider implementing a full partial information decomposition (e.g. as proposed in \parencite{bertschinger2014quantifying,james2018unique,rosas2020operational}) where redundancy and synergy can be considered as distinct information atoms that can be assessed entirely separately.

Another important open question is how to best visualize hypergraphs. It is known how pairwise networks can be elegant and visually inspiring, but when overcrowded  they can become hard to interpret --- resembling an spaghetti ball. 
In hypergraphs this visualization-level curse of dimensionality appears earlier, and we acknowledge that some of the plots presented here (made possible thanks to HyperNetX~\parencite{Praggastis2019HyperNetX}) can give inspiration for crop circles design. This is not a valid reason to escape the fact that networks are complicated and overcrowded, and hopefully new designs and specialised software will come up.

To conclude, by taking inspiration in complex system whose behavior is not equal to the one resulting from sum of its parts, here we asked ourselves whether the interdependencies observed in psychometric data could be described in a way that could account for beyond-pairwise (albeit conditioned), higher order interactions. 
Our proposed framework builds hyperlinks directly from the multivariate distribution of the data, identifying multiplets with a dominantly redundant or synergistic role. 
These analyses may ultimately inform on \emph{which kind} of information is being shared by different behavioral variables or symptoms, and how these informational multiplets change in time. 
This higher-order perspective 
complements current network approaches, leading to ways of assessing the data that greatly differ from what can be attained by other methods such as factor analyses. 
By doing this, the proposed approach enriches the toolbox of psychometricians and opens promising avenues for future investigation.

\newpage
\section{Data and code availability}
The empirical data are found at these repositories \url{https://osf.io/73c4q/} and \url{https://osf.io/mj5wa/}.
The code for simulations, analyses, and plotting is described in this github repository \url{https://github.com/danielemarinazzo/O_info_psycnet}.

\section{Acknowledgements}
F.R. was supported by the Ad Astra Chandaria foundation.
M.S. acknowledges Luigi Lombardi for discussions.
R.C. was supported by the Mind, Brain and Reasoning PhD programme at the University of Milan.
S.S. was supported by MIUR project PRIN 2017WZFTZP, “Stochastic forecasting in complex systems.”

\newpage

\printbibliography
\appendix
\section{Simulation of synergy and redundancy with \textit{lavaan}}
\label{app:procedure}

From Eqs.~\eqref{eq:Oinfo}, \eqref{eq:TC}, and \eqref{eq:DTC}, the O-information between three variables $(X,Y,Z)$ can be calculated as
\begin{equation}
\Omega(X;Y;Z) = I(X;Y) - I(X;Y|Z).
\end{equation}
Moreover, if we assume $(X,Y,Z)$ follow a multivariate Gaussian distribution, then their O-information can be calculated from their correlation matrix as follows:
\begin{equation}
\Omega(X;Y;Z) 
= \frac{1}{2} \log\frac{1-\rho(X,Y|Z)^2}{1 - \rho(X,Y)^2}
= \frac{1}{2}\log \frac{|\Sigma|}{|\Sigma_{xy}\Sigma_{yz}\Sigma_{xz}|},
\end{equation}
Above, the first equality uses the fact that for multivariate Gaussian variables $I(X;Y) = -\frac{1}{2} \log (1 - \rho(X,Y)^2)$ and $I(X;Y|Z) = -\frac{1}{2} \log (1 - \rho(X,Y|Z)^2)$, where $\rho(X,Y)$ is the Pearson correlation between $X$ and $Y$ and $\rho(X,Y|Z)$ is the conditional Pearson correlation of $X$ and $Y$ given $Z$, which can be calculated as
\begin{equation}
\rho(X,Y|Z) = \frac{\rho(X,Y) - \rho(X,Z) \rho(Y,Z)}
                   {\sqrt{1 - \rho(X,Z)^2}
				    \sqrt{1 - \rho(Y,Z)^2}}.
\end{equation}
The second equality can be directly verified, and denotes as $|\Sigma|$ the determinant of the correlation matrix of $(X,Y,Z)$ and $|\Sigma_{ij}|$ the determinant of the submatrix of the corresponding variables. 

Overall, the second expression shows that the O-information is transformation invariant (i.e. it does not depends on the mean values or variances of its arguments but only on the three correlations $\rho_{xy}$, $\rho_{yz}$, and $\rho_{xz}$), and the latter that it is symmetric on its three arguments.

While finding the interaction information that corresponds to three correlation values is straightforward, the reverse operation information is highly non-trivial. To facilitate this
process, we will use a parameterization based on factor analysis. The
factor model with a single factor and $n$ indicators $y_k$, $k=1,2,\ldots,n$,
can we written as
\begin{equation}
y_k = \lambda_k \eta + \nu_k
\end{equation}
where $\eta$ is a latent random variable, $\lambda_k\in\mathbb{R}$ is the factor
loading for indicator $k$, and $\nu$ is additive Gaussian noise with covariance matrix $\Theta$ with components $\theta_{i,j}$. 
For simplicity we set $\text{Var}(\eta) = 1$, $\text{E}(\theta_k) =  0$, and
$\text{Cov}(\eta, \theta_k) = 0$ for all $k=1,\dots,n$. 
Then it follows that the covariance matrix of $y_1,\dots,y_n$, denoted by $\Sigma$, is given by
\begin{equation}
\Sigma = \lambda \lambda^{T} + \Theta,
\end{equation}
where $\lambda=(\lambda_1,\dots,\lambda_n)$.

\begin{figure}[htp]
\centering
\includegraphics[width=.3\textwidth]{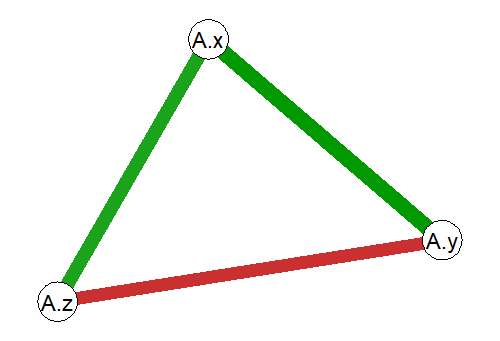}\hfill
\includegraphics[width=.3\textwidth]{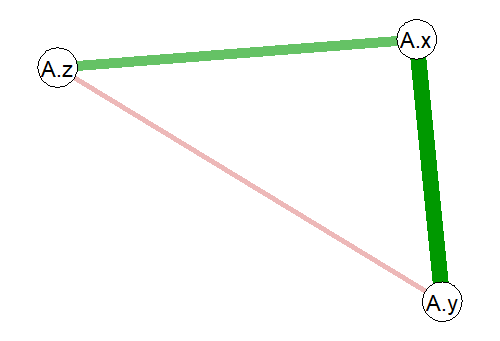}\hfill
\includegraphics[width=.3\textwidth]{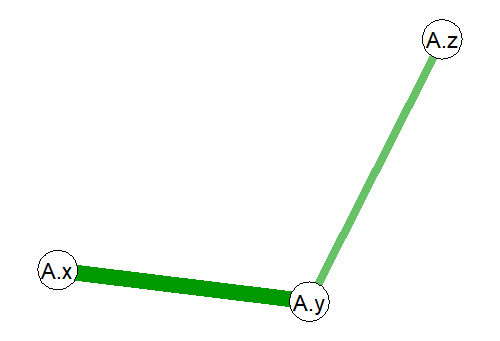}
\caption{\textit{qgraph} visualizations of the triplet. Both color intensity
and edge thickness represent edge weight. Red indicates negative edge weight,
green indicates positive edge weight. Left: synergy; Middle: Zero interaction
information; Right: redundancy.}
\label{qgraph_examples}
\end{figure}

If we have a triplet of three variables $X$, $Y$ and $Z$, the
factor model can be expressed by the equations:
\begin{align*}
X = \lambda_x \eta + \theta_x\\
Y = \lambda_y \eta + \theta_y\\
Z = \lambda_z \eta + \theta_z.
\end{align*}
In order to create a correlation matrix which either exhibits redundancy,
independence or synergy, we will first fix the factor loadings to apriori
chosen constants. Several choices are possible, but we have used $\lambda_x =
\sqrt{0.99}$, $\lambda_y = \sqrt{0.70}$ and $\lambda_z = \sqrt{0.30}$. To
ensure that the resulting covariance matrix has a unit diagonal, we set the
value of $\theta_x$, $\theta_y$ and $\theta_z$ to $(1 - 0.99)$, $(1 - 0.70)$
and $(1 - 0.30)$ respectively. The only parameter that we will vary is the
error covariance (ecov) $\text{Cov}(\theta_y, \theta_z)$. All other
off-diagonal elements of $\bm{\Theta}$ are fixed to zero.  The possible range
of this ecov parameter is about $[-0.458, +0.458]$ as determined by $\sqrt{1 -
0.70} \times \sqrt{1 - 0.30}$.  Otherwise, the correlation matrix would no
longer be positive definite. Changing this ecov parameter allows us to set the
level of interaction information. For example, if we let ecov to be $-0.14849$,
the level of interaction information is smaller than 1e-07, implying
independence. If we let ecov to be $-0.39$ or $0.22$, the level of information
is about $0.58$ (indicating synergy) and $-0.175$ (indicating redundancy)
respectively.  The corresponding correlation matrices are shown in
Table~\ref{covariance_matrix_zero}, Table~\ref{covariance_matrix_synergy} and
Table~\ref{covariance_matrix_redundancy}, where we use $A.x$, $A.y$ and $A.z$
to refer to the three variables that make up the triplet $A$.
\begin{table}[!ht]
\small
\centering
\makebox[0pt][c]{\parbox{1\textwidth}{%
    \begin{minipage}[h]{0.32\hsize}
        \begin{tabular}{l|lll}
        & A.x   & A.y   & A.z  \\ \cline{1-4}
        A.x & 1     &      &   \\
        A.y & 0.832 & 1     &   \\
        A.z & 0.545 & \textbf{0.310} & 1
        \end{tabular}
        \vspace*{1cm}
        \caption{Zero interaction, with $\textit{ecov} = -0.14849$}
        \label{covariance_matrix_zero}
    \end{minipage}
    \hfill
    \begin{minipage}[h]{0.32\hsize}
        \begin{tabular}{l|lll}
        & A.x   & A.y   & A.z \\ \cline{1-4}
        A.x & 1     &      &  \\
        A.y & 0.832 & 1     & \\
        A.z & 0.545 & \textbf{0.068} & 1
        \end{tabular}
        \vspace*{1cm}
        \caption{Synergy, with\\ $\textit{ecov} = -0.39$}
        \label{covariance_matrix_synergy}
    \end{minipage}
    \hfill
    \begin{minipage}[h]{0.32\hsize}
        \begin{tabular}{l|lll}
        & A.x   & A.y   & A.z \\ \cline{1-4}
        A.x & 1     &      &  \\
        A.y & 0.832 & 1     & \\
        A.z & 0.545 & \textbf{0.678} & 1
        \end{tabular}
        \vspace*{1cm}
        \caption{Redundancy, with $\textit{ecov} = 0.22$}
        \label{covariance_matrix_redundancy}
    \end{minipage}
}}
\end{table}
Obviously, the level of interaction information will have an impact
on the partial correlations. As a result, if we apply the graphical
lasso to these triplets and visualize them using the \textit{qgraph}
package, they result in very different patterns as can be
seen in Figure~\ref{qgraph_examples}.

\begin{figure}[ht]%
\includegraphics[scale=0.6]{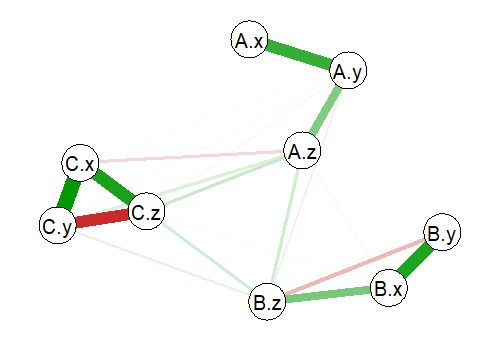}%
\centering
\caption{\textit{qgraph} network visualization of three triplets with residual
correlations between triplets. Triplet A includes redundancy; triplet B
includes zero interaction information; triplet C includes synergy. The primary
link between the triplets are done via the variables \textit{A.z}, \textit{B.z}
and \textit{C.z}. Residual correlations were set at 0.15.}%
\label{cluster_example}%
\end{figure}

A final model can combine several triplets, each with its own value for the
ecov parameter. The full covariance matrix is then constructed by placing the
model-implied covariance (correlation) matrices of the triplets on the main
diagonal. In addition, we can allow for variables to be connected by setting
the corresponding off-diagonal elements of the covariance matrix to a non-zero
value. An example covariance matrix for three connected triplets is shown in
Table~\ref{covariance_matrix_cluster}.
\begin{table}[ht]
\centering
\begin{tabular}{l|lllllllll}
 & A.x  & A.y  & A.z  & B.x  & B.y  & B.z  & C.x & C.y  & C.z \\ \hline
A.x & 1 &   &   &   &   &   &  &   &  \\
A.y & \textit{0.832} & 1 &   &   &   &   &  &   &  \\
A.z & \textit{0.545} & \textit{0.678} & 1 &   &   &   &  &   &  \\
B.x & 0 & 0 & 0 & 1 &   &   &  &   &  \\
B.y & 0 & 0 & 0 & \textit{0.832} & 1 &   &  &   &  \\
B.z & 0 & 0 & \textbf{0.150} & \textit{0.545} & \textit{0.308} & 1 &  &  & \\
C.x & 0 & 0 & 0 & 0 & 0 & 0 & 1   &  & \\
C.y & 0 & 0 & 0 & 0 & 0 & 0 & \textit{0.832}& 1 & \\
C.z & 0 & 0 & \textbf{0.150} & 0 & 0 & \textbf{0.150} & \textit{0.545}& \textit{0.068} & 1
\end{tabular}
\caption{Covariance matrix representation of the example displayed in Figure
\ref{cluster_example} including off-diagonal covariances in bold.}
\label{covariance_matrix_cluster}
\end{table}
A visualization of this covariance matrix is presented in 
Figure~\ref{cluster_example}

\appendix
\section{Reanalysis of the Golino \textit{redundancy} framework}
\label{app:Golino_red}

\begin{figure}[!ht]%
\includegraphics[width=.6\textwidth]{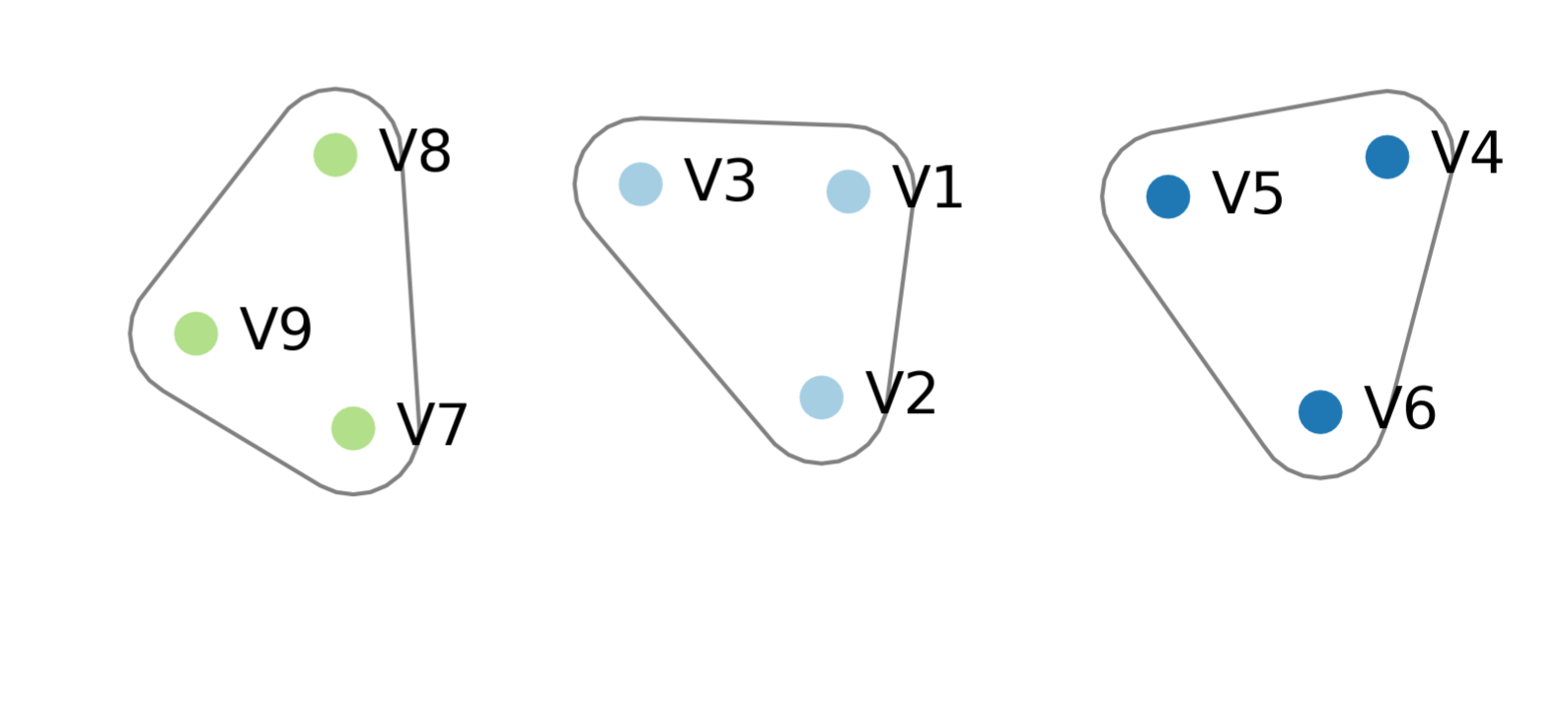}%
\centering
\caption{Significant redundant multiplets when the factors are not correlated.}%
\label{fig_Golino_0}%
\end{figure}

\begin{figure}[!ht]%
\includegraphics[width=\textwidth]{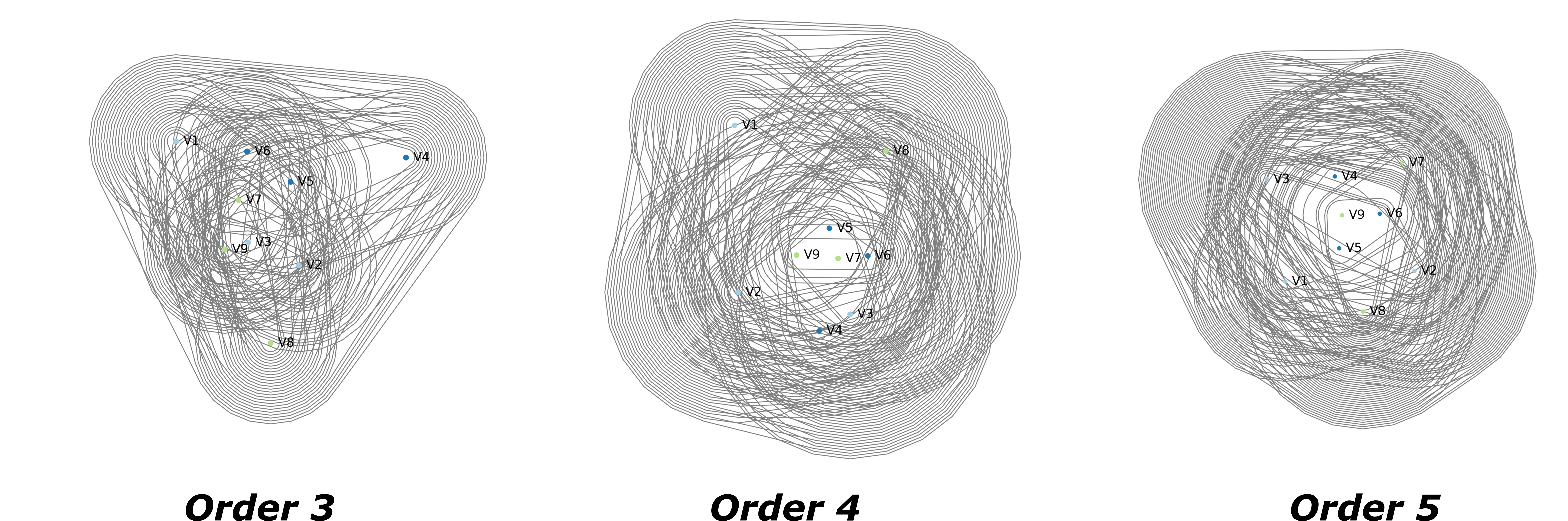}%
\centering
\caption{Significant redundant multiplets when the factors are correlated.}%
\label{fig_Golino_1}%
\end{figure}

Here we analyze the datasets simulated in~\parencite{christensen2020unique,golino2021entropy,golino2021modeling} with 2000 observations, three latent factors (factor loadings $= 0.4$ and three variables per latent factor, without ($c=0$, \ref{fig_Golino_0}) and with ($c=0.3$, \ref{fig_Golino_1}) correlation between factors. Only redundancy emerges from the analysis with our framework.

\end{document}